\documentstyle[aps]{revtex}  
\begin{document} 
\def\Astrobiologia{Astrobiolog\'\i{}a}
\def\Astrofisica{Astrof\'\i{}sica}
\def\Paris{Par\'\i{}s}
\def\Perez{P\'erez}
\def\Fisica{F\'\i{}sica}
\def\Torrejon{Torrej\'on}
\title{
Heat kernel regularization of the effective action for stochastic
reaction--diffusion equations}
\author{ 
David Hochberg,$^{+,++,a}$ 
Carmen Molina--\Paris,$^{++,+++,b}$ 
and Matt Visser$^{++++,c}$ 
} 
\address{ 
$^{+}$Laboratorio de {\Astrofisica} Espacial y {\Fisica} 
Fundamental, Apartado 50727, 28080 Madrid, Spain\\ 
$^{++}$Centro de \Astrobiologia,
CSIC-INTA, Carretera de Ajalvir Km. 4, 28850 
{\Torrejon} de Ardoz, Madrid, Spain\\
$^{+++}$Theoretical Division T-8, Los Alamos National Laboratory, Los
Alamos, New Mexico 87545\\ 
$^{++++}$Physics Department, Washington University, 
Saint Louis, Missouri 63130-4899, USA} 
\date{\today} 
\maketitle 
\begin{abstract}
The presence of fluctuations and non-linear interactions can lead to
scale dependence in the parameters appearing in stochastic
differential equations.  Stochastic dynamics can be formulated in
terms of functional integrals.  In this paper we apply the heat kernel
method to study the short distance renormalizability of a stochastic
(polynomial) reaction-diffusion equation with real additive noise.  We
calculate the one-loop {\emph{effective action}} and its ultraviolet
scale dependent divergences.  We show that for white noise a
polynomial reaction-diffusion equation is one-loop {\emph{finite}} in
$d=0$ and $d=1$, and is one-loop renormalizable in $d=2$ and $d=3$
space dimensions. We obtain the one-loop renormalization group
equations and find they run with scale only in $d=2$.

{PACS number(s): 02.50.Ey;02.50.-r;05.40.+j}

{\emph{Keywords}}:  Reaction-diffusion; Effective action; Beta functions
\end{abstract}
\def\Leon{Le\'on}
\def\Quiros{Quir\'os}
\def\Barabasi{Barab\'asi}
\def\Dominguez{Dom\'{\i}nguez}
\def\ie{\emph{i.e.}}
\def\eg{\emph{e.g.}}
\def\d{{\mathrm{d}}}
\def\half{{1\over 2}}
\def\A{{\mathcal A}}
\def\J{{\mathcal J}}
\def\field{{\mathrm{field}}}
\def\ghost{{\mathrm{ghost}}}
\def\free{{\mathrm{free}}}
\def\finite{{\mathrm{finite}}}
\def\oneloop{{\mathrm{1-loop}}}
\def\cut-off{{\mathrm{cut-off}}}
\def\HRULE{ \bigskip \hrule \bigskip}
\newcommand{\define}{\mathop{\stackrel{\rm def}{=}}}
\newcommand{\tr}{\mathop{\mathrm{tr}}} 
\newcommand{\Tr}{\mathop{\mathrm{Tr}}}
\def\e{\epsilon}

\section{Introduction}

Many examples abound where particular spacetime distributions of
matter are selected over a wide variety of seemingly possible
choices. In many of these cases, these patterns are well described,
and their temporal evolution accurately modelled, by non-linear
partial differential equations subject to noise, or equivalently, by
stochastic partial differential equations (SPDEs).  Concrete examples
can be found in the domains of pattern formation, chemical chaos,
biological morphogenesis, and flame-front propagation, just to name a
few~\cite{Cross-Hohenberg,Walgraef,Sivashinski}.  As argued in
Ref.~\cite{priority2a} the {\emph{effective potential}} is a superb
tool for studying the onset of pattern formation (\ie, symmetry
breaking) about the static and spatially homogeneous solutions of the
SPDE.  This quantity takes into account both interactions and
fluctuations to a given number of {}\emph{loops}. (The loop expansion
is a controlled expansion in the amplitude of the fluctuations.) To go
beyond the study of symmetry breaking and static homogeneous
configurations, it is crucial to be able to include and account for
time development and spatial inhomogeneities.  Ultimately, we are
interested in studying the late time behavior of the solutions of
SPDEs; we would like to know if there is an asymptotic steady state,
or state of equilibrium, as this impinges on the late time behavior of
the emergent pattern.  The effective potential does not yield this
information, and one must turn to the {\emph{effective action}}.  The
effective action contains all the dynamical solutions of the SPDE and
their asymptotic behavior. It obeys a variational principle wherein
its first variation yields the SPDE. Its exact calculation, however,
is very difficult. Along the way, one must first ``regularize'' its
short distance divergences that are dimension dependent.  These must
be identified, isolated, regulated, and if possible, renormalized by
suitable redefinitions of the bare parameters appearing in the
SPDE. The study of these dimension dependent divergences lends itself
to a direct calculation of the renormalization group equations (RGEs)
that encode the scale dependence (or ``running'') of these
parameters. The RGE fixed points provide information about the
asymptotic states. For these and other reasons, the divergent
structure of the effective action warrants a detailed analysis in its
own right.

In this paper we study the stochastic reaction-diffusion (RD) equation
by means of a functional integral representation that will be used to
calculate its one-loop effective action.  In earlier work we
calculated and analyzed the {\emph{effective potential}} for the RD
equation (the effective action evaluated for special field
configurations that are time independent)~\cite{priority2c}. However,
the effective potential does not provide information about the
dynamics of the system or wavefunction renormalization.  The effective
potential can only signal the static homogeneous states around which
one can study the onset of pattern formation (\ie, symmetry breaking),
but does not indicate {\emph{which}} such state (if any) is
dynamically accessible or most likely.  To go beyond the limitations
inherent in an effective potential analysis, and in order to judge the
importance of the noise-induced symmetry broken ground state
configuration, one should explore the space of dynamical (spatially
inhomogeneous) solutions of the system.  For instance, are there
solutions that indicate the system is ``thermalizing'', or can we find
steady state solutions?  An important step in this direction is the
analysis of the effective action undertaken in this paper.  In this
calculation we encounter short distance divergences that need to be
regularized.  We have chosen to carry out this regularization
procedure by means of (generalized) heat kernel techniques.  To do so,
we will be required to calculate the first (integrated) Seeley--DeWitt
coefficients.  In addition to the physical RD field itself, we must
also calculate the contribution from the ghost field, a necessary
ingredient in this formalism~\cite{priority2a}.

After the regularization step, we turn to the one-loop renormalization
of the effective action.  We find that for additive white noise RD
systems are one-loop finite in $d=0$ and $d=1$ space dimensions, and
(at least) one-loop renormalizable in $d=2$ and $d=3$ space dimensions
for polynomial reaction kinetics. There is no wavefunction
renormalization in any of these dimensions, irrespective of the degree
of the reaction polynomial~\cite{Cardy}.  Moreover, the ultraviolet
renormalizability requires including a bare tadpole $\lambda_0$, or
constant term, in the equation of motion.

Application of heat kernel methods to stochastic field theories far
from equilibrium may not be familiar, so a few comments regarding them
may be useful.  These techniques have been used primarily for
calculating effective actions and the one-loop physics in quantum
field theories (QFTs) and in curved space quantum field
theory~{\cite{Birrell,Fulling,Grib,DeWitt,Gibbons,Bordag}}.  In the
quantum domain, one is interested in computing quantities such as the
{\emph{effective action}} and the {\emph{effective potential}}, which
provide crucial information regarding the structure of the underlying
theory at different length and time scales and are important in
assessing the theory's renormalizability (or lack thereof), the
determination of the running of couplings and parameters, patterns of
spontaneous and dynamical symmetry breaking, and the structure of
short distance (ultraviolet) and long distance (infrared)
divergences~{\cite{QFT,Ramond,Zinn,Rivers,Weinberg}}.  Moreover, for
renormalizable theories, the computation of the effective action
(actually, only its divergent part is needed) can be used to extract
the RGEs that govern the scale dependence of the couplings and
parameters appearing in the theory~\cite{Fujimoto,Gato,phi4}.  Though
perhaps better known in the context of these fields, these same
techniques can be generalized and applied to reveal the corresponding
one-loop physics associated with stochastic dynamic phenomena and to
systems subject to fluctuations.

As a key technical step, we need to obtain the Green functions for
certain higher-order differential operators that are neither elliptic
nor hyperbolic. (The operators treated here are second-order in
physical time derivatives and fourth-order in space derivatives.) We
set up a generalization of the Schwinger proper time asymptotic
expansion to compute these Green functions and obtain explicit
expressions for the first Seeley--DeWitt coefficients associated with
these operators.  We carry this out employing the minimal
representation in which only physical and ghost field variables appear
in the action~\cite{priority2a}.

At this stage we emphasize the following: We start from a given
``classical'' SPDE, together with a specification of the type of noise
and its correlations. Our aim is to {\emph{map}} this SPDE to an
equivalent generating functional and effective action, at which point
QFT techniques can be used.  We emphasize that our approach and use of
field theory is rather different from that initiated some years ago,
whose aim is to {\emph{derive}} the SPDE and the noise correlations
from microphysics by mapping a classical master equation to
``second-quantized'' variables, and then finally to an action
principle~\cite{Doi,Peliti,Grassberger}.

We conclude with a summary and discussion of this work, and prospects
for further development.

\section{Stochastic field theory for reaction-diffusion equations}

In this paper we demonstrate that heat kernel techniques can be used
to compute the ultraviolet divergent terms arising in the one-loop
effective action associated with field theory formulations of
SPDEs. In particular, we apply these methods to the class of single
component reaction-diffusion equations
\begin{equation}
\label{E:RDD}
D\phi(x) = V[\phi(x)] + \eta(x),
\end{equation}
where
\begin{equation}
\label{E:diffusion-operator}
D = \frac{\partial}{\partial t} - \nu \nabla^2,
\end{equation}
and
\begin{equation}
\label{E:potential-term}
V[\phi] = \sum_{j=0}^{N} \frac{\lambda_j}{j!}\phi^j(x).
\end{equation}
Here $\eta(x)$ is a noise term with normalized probability
distribution given by $P[\eta]$. We employ the shorthand notation $x =
(\vec x,t)$.  Furthermore, $V[\phi]$ is a polynomial of degree $N$ in
the concentration (or field variable $\phi$) and the $\lambda_j$'s are
a set of reaction rates.  For convenience we have placed the decay (or
``mass'' term) into the potential and have treated it as another
coupling: $\lambda_1 \equiv - m^2$.  It must be noted that this
equation has the form of a purely dissipative system and has a
bona-fide potential energy term $V[\phi]$~\cite{Zinn}.  Thus, it makes
sense to calculate an {}\emph{effective potential} for constant field
configurations, as well as an effective action for inhomogeneous
fields.  Both the effective action and the effective potential are
derived by means of a field theory for this SPDE.

Before continuing, we should point out that there are reasons to
believe that as a {}\emph{phenomenological} equation, the RD being
considered here, (and others structurally similar to it), might not be
adequate to describe certain two-body annihilation processes, or pair
reaction kinetics, since recent derivations based on master equations
show that the SPDE in question should actually be {}\emph{complex}
with {}\emph{imaginary} noise (leading to negative noise
correlations)~\cite{Cardy,Howard,Rey-Cardy}.  On the other hand, for
processes involving particle clustering, these same derivations yield
real stochastic equations and noise, as well as positive noise
correlations~\cite{real-noise}.  Of course, there are many situations
in which a microscopic derivation of the SPDE is entirely out of the
question, either because explicit knowledge of the microscopic details
is lacking and/or because the random fluctuations owe to
uncontrollable contingencies. In these situations the benefit of
adopting a phenomenological strategy should be self-evident. Finally,
the application of this equation need not be restricted to just
chemical diffusion~\cite{Murray}.

For homogeneous and static concentrations it is sufficient to study
the effective potential~\cite{priority2c}.  In this paper we
complement that analysis by making use of the effective action to
consider inhomogeneous and time dependent field configurations.  In
the minimalist approach (see Ref.~\cite{priority2a}) one starts with
the normalized generating functional $Z[J]$ encoding the stochastic
dynamics described by the RD equation (\ref{E:RDD}). This involves the
RD scalar field $\phi$ plus the Jacobian determinant (denoted here by
$\cal J$) and its adjoint ($\cal J^\dagger$), (these determinants
arise from a change of variables). The generating functional
(partition function) is given by~\cite{priority2a,Zinn}
\begin{equation}
\label{E:partition-function}
Z[J] = \frac{\int [\d\phi]\; 
\exp \{ [-S[\phi] + \int \d^n x \;J(x)\phi(x) ]/\A \} \;
\sqrt{{\cal J}{\cal J}^\dagger} }
{\int [\d\phi]\; \exp \{ -S[\phi]/\A  \} \; 
\sqrt{{\cal J}{\cal J}^\dagger}  },
\end{equation}
where the ``classical'' action is [valid for Gaussian noise:
$G_{\eta}(x,y) = \langle \eta(x) \eta(y) \rangle = \A\; g_2(x,y)$]
\begin{equation}
\label{E:action1}
S[\phi] = \half\int \d^nx_1 \int \d^nx_2\;
(D\phi(x_1) -V[\phi(x_1)]) \;
g_2^{-1}(x_1,x_2) \;
(D\phi(x_2) -V[\phi(x_2)]),
\end{equation}
with $n=d+1$ and $d$ the number of space dimensions (we will keep $d$
as a free parameter throughout this paper). For a general SPDE there
may be non-vanishing contributions from the ``ghost'' fields (Jacobian
determinants).  We follow here the discussion of
Ref.~\cite{priority2a} to separate the noise two-point function into
the product of a {}\emph{shape} $g_2(x,y)$ and a constant amplitude
${\cal A}$.  Irrespective of how we decide to normalize the shape, the
constant amplitude $\A$ is always the {}\emph{loop-counting} parameter
of the perturbation expansion~\cite{priority2a}.  A loop-counting
parameter is very useful in organizing such a perturbative expansion.
Moreover, any symmetry that is present in the classical action
(\ref{E:action1}) is preserved at each order in the loop expansion
since the loop-counting parameter multiplies the entire action (and
the source term $J$) in (\ref{E:partition-function}).  One of the
advantages of the minimal representation is that it leads to this
natural identification of the noise amplitude~\cite{priority2a,Zinn}.

We introduce the generating functional for connected correlation
functions $W[J]$ and its Legendre transform, the effective action
$\Gamma[\bar \phi]$~\cite{Zinn} , (note the explicit factor of the
noise amplitude $\A$)
\begin{mathletters}
\begin{eqnarray}
\label{E:Legendre}
W[J] &=& + \A \; \log Z[J],
\\
\Gamma[\bar \phi] &=& -W[J] + 
\int \d^nx \; J(x)\left\{ \bar \phi[J](x) - \bar\phi[0](x) \right\},
\end{eqnarray}
\end{mathletters}
with
\begin{equation}
\bar \phi [J] = \frac{\delta W [J]}{\delta J}.
\end{equation}
The barred fields $\bar \phi[J]$ and $\bar \phi[0]$ are the solutions
of the equations of motion
\begin{equation}
\label{E:classical-eom}
\left.\left(\frac{\delta \Gamma[\bar \phi]}{\delta \bar \phi}
\right)\right|_{\bar \phi [J]} = J(x), 
\qquad {\rm and} \qquad
\left.\left(\frac{\delta \Gamma[\bar \phi]}{\delta \bar \phi}
\right)\right|_{\bar \phi [0]} = 0,
\end{equation}
respectively. It is usually assumed that the former equation has a
unique solution $\bar \phi[J]$ (at least for small $J$), and that for
vanishing source ($J=0$) the unique solution is the vanishing mean
field $\bar \phi[0] = 0$, \ie, $\langle \phi(\vec x,t) \rangle_{J=0} =
0$, where the angular brackets denote the stochastic average with
respect to the noise probability distribution ${\cal P}[\eta]$.  This
is valid for a symmetric ground state.  We next expand the action
(\ref{E:action1}) about this background up to quadratic order in the
stochastic fluctuation $\delta \phi = \phi - \bar \phi$.  We can carry
out a perturbation expansion in the small parameter $\A$ and compute
the one-loop effective action to obtain~\cite{QFT}~\footnote{We now
drop the overbar on $\phi$ with the understanding that this stands for
the conjugate field of $J$ and not the field appearing in the
classical (zero-loop) action.}
\begin{eqnarray}
\label{E:effective1}
\Gamma[\phi] &=& 
S [\phi] -S[0] 
+ \frac\A{2}\left(\Tr \log S^{(2)}_\field[\phi]
-\Tr \log S^{(2)}_\field[0]  
-\log {\cal J}[\phi] - \log {\cal J}^\dagger[\phi]
+\log {\cal J}[0] +\log {\cal J}^\dagger[0]
\right)
+O(\A^2)
\nonumber \\
&=&  
S [\phi] - S[0] 
+ \Gamma^{(1-loop)}[\phi] 
+O(\A^2)
\nonumber \\
&=&  
S [\phi] - S[0] + \Gamma_{\epsilon}^{(1-loop)}[\phi] +
\Gamma_{finite}^{(1-loop)}[\phi]
+O(\A^2),
\end{eqnarray}
where the matrix elements of the Jacobi field operator
$S^{(2)}_\field[\phi]$
are
\begin{equation}
\label{E:Jacobi}
\langle x_1|S^{(2)}_\field[\phi]| x_2 \rangle =
S^{(2)}_\field(\phi,x_1,x_2) =
\frac{\delta^2 S[\phi]}{
\delta \phi(x_1) \; \delta \phi(x_2)}.
\end{equation}
We have anticipated the appearance of divergences in the one-loop
contribution to the effective action, arising from both the physical
and ghost fields, and have supplied it with a cut-off $\epsilon$.

The notation $S[0]$ is actually shorthand for $S[\phi[J=0]]$, and for
a symmetric ground state one typically has $\phi[0] = 0$ and $S[0] =
0$, unless there is a ``tadpole'' in the classical action. In fact,
when looking for mean field solutions of the zero-loop equation of
motion, we will find it convenient to consider a non-vanishing value
of the mean field $\phi[0] = v_0 \neq 0$ and will study fluctuations
about this mean value.  The terms involving $S[0]$ and
$S^{(2)}_\field[0]$ appear due to the normalization factor in
(\ref{E:partition-function}).  The notation ``$\rm Tr$'' stands for
the trace and indicates that we are to take the (time and space)
coincidence limits $x_2 \rightarrow x$ and $x_1 \rightarrow x$,
followed by an integration over the common limit $x$. The one-loop
effective action will contain divergent terms and it is precisely
these terms we wish to isolate and compute.  We have collected all
such divergences into the expression $\Gamma_{\epsilon}^{(1-loop)}$
and we regulate them by means of a cut-off $\epsilon$.  The finite
terms are collectively represented by $\Gamma_{finite}^{(1-loop)}$.
There may also be higher-loop contributions, denoted by $O(\A^2)$,
whenever we need to emphasize them explicitly.  Although these latter
contributions are important for constructing the full effective
action, that calculation is beyond the scope of the present paper.

In order to compute the one-loop effective action we need to obtain
$S^{(2)}_\field[\phi]$.  This Jacobi field operator is diagonal in
coordinate space.  For the purpose of this calculation and in the
interest of simplicity, we consider the case of white noise. (Colored
noise can be dealt with, but it brings in time and space derivatives
of the shape function, which complicate the heat kernel analysis.) For
white noise we have $\langle \eta(x) \eta(y) \rangle = 2D_0\;\delta^n
(x,y)$ and therefore we can write
\begin{eqnarray}
\label{E:noise}
G_{\eta}(x,y) &=& 2 \, D_0 \; \delta^n(x,y),
\; \; \; 
\Rightarrow \A  = 2 \, D_0, 
\; \; \; 
 g_2(x,y) = \delta^n(x,y),
\; \; \; 
{\rm and}
\; \; \; 
 g_2^{-1}(x,y) = \delta^n(x,y),
\end{eqnarray}
which fixes the noise normalization. 

The Jacobi field operator corresponding to the RD equation is easy to
calculate starting from the classical action. We simplify notation and
write the zero-noise action as
\begin{eqnarray}
\label{E:action3}
S[\phi] &=& \half \int \d^nx\; \left(D\phi - V[\phi] \right)^2
= \half \int \d^nx\; \left(\partial_t \phi -\nu \nabla^2
\phi - V[\phi] \right)^2.
\end{eqnarray}
The Jacobi operator for the physical field,
$S^{(2)}_\field(\phi,x_1,x_2)$, is given by
\begin{eqnarray}
\label{E:jacobi-field}
S^{(2)}_\field(\phi,x_1,x_2) 
&\equiv&
\left[
(-\partial_t -\nu \nabla^2  - V'[\phi(x_1)])
( \partial_t - \nu \nabla^2 - V'[\phi(x_1)])   
-V''[\phi(x_1)] 
(D \phi(x_1) - V[\phi(x_1)])
\right]
\delta^n(x_1,x_2),
\end{eqnarray}
where $V'[\phi] = \frac{\d V[\phi]}{\d\phi}$ and  $V''[\phi] = \frac{\d^2
V[\phi]}{\d\phi^2}$.

For the ghost field the Jacobi operator is given by~\cite{priority2a}
\begin{equation}
\label{E:jacobi-ghost}
S^{(2)}_\ghost(\phi,x_1,x_2) 
\equiv
\Big\{ 
(-\partial_t -\nu \nabla^2  - V'[\phi(x_1)])
( \partial_t - \nu \nabla^2 - V'[\phi(x_1)])
\Big\}
\delta^n(x_1,x_2),
\end{equation}
and its determinant can be written as~\cite{priority2a}
\begin{equation}
\det[S^{(2)}_\ghost(\phi,x_1,x_2) ] = 
{\cal J}[\phi] \;  {\cal J}^\dagger[\phi].
\end{equation}

In order to carry out the perturbation expansion we also need to
consider the ``free'' case defined by the limit $V[\phi]\to0$
\begin{eqnarray}
\label{E:jacobi-free}
S^{(2)}_\free(x_1,x_2) &=&
\left[
(-\partial_t -\nu \nabla^2 ) \;
( \partial_t -\nu \nabla^2 ) 
\right] \;
\delta^n(x_1,x_2)
= 
\Big[
-\partial_t^2 + ( \nu \nabla^2)^2
\Big] \;
\delta^n(x_1,x_2).
\end{eqnarray}
Free physical fields have the same Jacobi operator as free ghost
fields, so that as $V[\phi]\to0$, the physical and ghost field
contributions to the effective action cancel.  We now look ahead a
little: as the Jacobi operator $S^{(2)}_{\mathrm free}(x_1,x_2)$
contains fourth order space derivatives ({\emph{bi-harmonic
operator}}), rather than second order derivatives, the behavior of the
DeWitt--Schwinger
expansion~{\cite{Birrell,Fulling,Grib,DeWitt,Gibbons}} is
{\emph{qualitatively}} different in that it includes
{\emph{fractional}} powers of the Schwinger proper time parameter.

We now calculate the mean field $v_0$ (\ie, the background field) by
studying the solutions of the classical equation of motion, which is
given (for arbitrary source $J$) by
\begin{equation}
\label{E:zeroloop}
\left.\left(\frac{\delta S}{\delta \phi}\right)\right|_{\phi [J]}
=
(-\partial_t -\nu \nabla^2 - V'[\phi(x)]) \;
(D\phi(x) -V[\phi(x)]) 
= J(x).
\end{equation}
If the source vanishes and the mean field is homogeneous and static,
we have
\begin{equation}
\label{E:vev}
V'[v_0] \; V[v_0] = 0,
\end{equation}
which always has at least one real solution~\cite{priority2c}.
 
In order to calculate the one-loop effective action for RDs, one must
include the contribution from the ``ghost'' fields.  These ``ghost''
Jacobians are given by~\cite{priority2c}
\begin{equation}
\label{E:jacobians}
{\cal J} = {\rm det}\left(D - \frac{\delta V}{\delta \phi}\right)
,
\; \; \; 
{\rm and}
\; \; \; 
{\cal J}^{\dagger} = {\rm det}\left(D^{\dagger} - 
\frac{\delta V^{\dagger}}{\delta \phi}\right).
\end{equation}
We can now complete the formal calculation of the one-loop effective
action.  We have~\cite{priority2a}
\begin{equation}
\label{E:effective2}
\Gamma[\phi] 
=
S[\phi] - S[v_0] + \frac\A{2}
\left[ 
\Tr \log S^{(2)}_\field[\phi] - 
\Tr \log 
\left(D^{\dagger} - \frac{\delta V^{\dagger}}{\delta \phi}\right)
\left(D - \frac{\delta V}{\delta \phi}\right) - (\phi \rightarrow v_0)
\right]
+O(\A^2),
\end{equation}
so the one-loop effective action receives one contribution from the
physical field
\begin{equation}
\Gamma^{(1-loop)}_\field = \frac\A{2}
\left(\Tr \log S^{(2)}_\field[\phi] -  \Tr \log S^{(2)}_\field[v_0] \right),
\end{equation}
and a contribution from the ghost field
\begin{equation}
\Gamma^{(1-loop)}_\ghost 
= 
- \frac{\A}{2} 
\left[
\Tr \log 
\left(D^{\dagger} - \frac{\delta V^{\dagger}}{\delta \phi}\right)
\left(D - \frac{\delta V}{\delta \phi}\right)_{\phi}
- 
\Tr \log 
\left(D^{\dagger} - \frac{\delta V^{\dagger}}{\delta \phi}\right)
\left(D - \frac{\delta V}{\delta \phi}\right)_{v_0}
\right].
\end{equation}
We will soon see that individually, each contribution has complicated
Seeley--DeWitt coefficients, but when taken together, the physical
plus ghost sectors yield simple {}\emph{net} Seeley--DeWitt
coefficients.

\section{Computing the one-loop effective action}

In this section we construct the regulated expression
$\Gamma_{\epsilon}^{(1-loop)}[\phi]$ for the RD equation. We follow
closely the DeWitt--Schwinger (DS) proper time formalism to analyze
the ultraviolet
divergences~{\cite{Birrell,Fulling,Grib,DeWitt,Gibbons,Ramond,Zinn,Rivers}}.
(We have striven to keep this section self-contained.)

In this formalism the integral representation for
$\Gamma^{(1-loop)}[\phi]$, eq. (\ref{E:effective1}), involves a
fictitious ``time'' parameter $s$ (denoted as Schwinger proper time).
To this end, we define the following function, where $A$ is any
operator
\begin{equation}
g_{\alpha}(A) \equiv \int_0^{+\infty} \d s \; s^{\alpha -1} e^{-sA}
= A^{-\alpha} \; \Gamma(\alpha),
\end{equation}
with $\Gamma(\alpha)$ the Gamma function.  We consider the limit
$\alpha \rightarrow 0$
\begin{equation}
g_{\alpha}(A) \rightarrow \frac{1}{\alpha} - \gamma - \log A,
\end{equation}
where $\gamma = 0.577...$, is Euler's constant. Although this integral
is divergent, the difference of {}\emph{two} such integrals is finite
and well defined
\begin{equation}
\lim_{\alpha \rightarrow 0}[g_{\alpha}(B) - g_{\alpha}(A)] = 
\log A - \log B = -\int_0^{+\infty} \frac{\d s}{s} (e^{-sA}- e^{-sB}),
\end{equation}
and comparing with (\ref{E:effective1}), the desired proper time
integral for the one-loop effective action is given by
\begin{eqnarray}
\label{E:representation}
\Gamma^{(1-loop)}[\phi] 
&=&
\Gamma^{(1-loop)}_\field[\phi]+\Gamma^{(1-loop)}_\ghost[\phi]
\\
&=&
-\frac\A{2}\int_0^{+\infty} \frac{\d s}{s}
\Tr \left( e^{-sH_\field} - e^{-s{[H_0]}_\field} \right)
+\frac\A{2}\int_0^{+\infty} \frac{\d s}{s}
\Tr \left( e^{-sH_\ghost} - e^{-s{[H_0]}_\ghost} \right),
\end{eqnarray}
where the ``Hamiltonians'' in the exponentials are the Jacobi
operators
\begin{mathletters}
\begin{eqnarray}
H_\field &=& S^{(2)}_\field[\phi[J]] = (D^\dagger - V')(D-V') - V''(D\phi -V), 
\\
{[H_0]}_\field &=& S^{(2)}_\field[\phi[0]], 
\\
H_\ghost &=& S^{(2)}_\ghost[\phi[J]] = (D^\dagger - V')(D-V') , 
\\
{[H_0]}_\ghost &=& S^{(2)}_\ghost[\phi[0]],
\\
H_\free &=& D^\dagger D,
\end{eqnarray}
\end{mathletters}
as can be seen by comparing (\ref{E:representation}) with
(\ref{E:effective1}), (\ref{E:jacobi-field}), (\ref{E:jacobi-ghost}),
and (\ref{E:jacobi-free}).  To proceed with the calculation, we need
an explicit form for the operators $e^{-sH}$ (that is, for
$e^{-sH_\field}$, $e^{-s{[H_0]}_\field}$, $e^{-sH_\ghost}$, and
$e^{-s{[H_0]}_\ghost}$), or rather, their matrix elements, so that we
can take the indicated traces. To solve for them, we note that
$e^{-sH}$ is the exact solution of the following operator differential
equation
\begin{equation}
\label{E:schrodinger} 
H e^{-sH} = -\frac{\partial}{\partial s} e^{-sH}.
\end{equation}
If one takes matrix elements in the spacetime coordinate basis $|x
\rangle \equiv |\vec x,t\rangle$, inserts a complete set of states,
and makes use of the diagonality of $H$ in the coordinate basis [note
that $S^{(2)}_{\rm field}[\phi]$ is proportional to
$\delta^n(x_1,x_2)$], we obtain
\begin{eqnarray}
\label{E:schrodinger2}
H(x) \; \langle x |e^{-sH}|x' \rangle  
&=& 
\int \d y \langle x |H| y \rangle \langle y| e^{-sH}|x' \rangle 
=
\langle x |H \; e^{-sH}|x' \rangle 
=-\frac{\partial}{\partial s} 
\langle x |e^{-sH}|x' \rangle,
\end{eqnarray}
or equivalently
\begin{equation} 
\label{E:schrodinger3}
H(x) \; G(x,x'|s) 
= 
-\frac{\partial}{\partial s} G(x,x'|s),
\; \; \; {\rm with}
\; \; \; G(x,x'|s) \equiv \langle x |e^{-sH}|x' \rangle.
\end{equation}
This latter equation defines the Green function $G(x,x'\vert s)$ in
terms of the matrix element of the operator $e^{-sH}$ in the
coordinate representation.  These steps can be repeated for the other
Hamiltonians.  Fortunately, for the purposes of the present work, it
is not necessary to solve this equation exactly (for either $H_\field$
or $H_\ghost$), as we are interested in the short distance divergent
part of the one-loop effective action.  What we will do instead is
solve the ``heat'' equations (\ref{E:schrodinger}),
(\ref{E:schrodinger2}), and (\ref{E:schrodinger3}) adiabatically by
expanding in small positive fractional powers of the proper time
variable $s$, which is where all the ultraviolet divergences are to be
found (different techniques are required if one is interested in the
infrared limit).  Nevertheless, to get ``off the ground'' it will be
most useful to have the exact solution to (\ref{E:schrodinger2}) in
the free limit ($V[\phi]\to0$). We now turn to this task, which
entails solving exactly (\ref{E:schrodinger}) with $H_\free$.

Since equation (\ref{E:schrodinger2}) looks like a heat equation in a
$n+1$ dimensional spacetime (parabolic equation in the proper time
variable), we know how to solve it together with specified boundary
and/or initial conditions.  In the free field limit ($G\rightarrow
G_\free$), we must solve the following equation
\begin{equation}
\label{E:freekernel-1}
\left[ -\partial_t^2 + (\nu \nabla^2)^2
 +
\frac{\partial}{\partial s} \right] G_\free(x,x'\vert s) 
= 
\delta^n(x-x') \; \delta(s),
\end{equation}
subject to the boundary (initial) condition $G_\free(x,x'\vert0) =
\delta^n(x-x')$.  Strictly speaking, the Green function depends on
both arguments $x$ and $x'$, but due to the translational invariance
of the dynamical equation, we have $G(x,x'|s) = G(x-x'|s) = G(\vec x -
\vec x', t - t'|s)$ and it suffices to treat it as a function of one
spacetime coordinate.  We can always restore its dependence on both
spacetime arguments at any time.

The formal solution for $G_\free$, expanded to fourth order in $x-x'$
(along the diagonal), is given by
\begin{eqnarray}
\label{E:formal}
G_\free(x,x'|s) &=&
A_d \; \exp\left[-{(t-t')^2\over4s}\right]\;  
s^{-{1\over2}-{d\over4}} 
\Bigg[ 
1 
- {C_{d,1}\over 2d} {|\vec x-\vec x'|^2 \over\nu\sqrt{s}}
+ {C_{d,2}\over8d(d+2)} {(|\vec x-\vec x'|^2)^2 \over\nu^2 \; s} 
+ O\left( {(|\vec x-\vec x'|^2)^3 \over s^{3/2}}\right)
\Bigg],
\end{eqnarray}
where 
\begin{equation}
A_d=\left(\frac{1}{4\pi}\right)^{\half} \;
\frac{\pi^{\frac{d}{2}} \; \Gamma(\frac{d}{4}) }
{2(2\pi)^d \; \Gamma({d\over2})}
\left( \frac{1}{\nu^2} \right)^{\frac{d}{4}},
\; \; \; 
{\rm and}
\; \; \; 
C_{d,n} = {\Gamma({d+2n\over4})\over\Gamma({d\over4})}.
\end{equation}
Details of the calculation leading to the final expression for
$G^{({\mathrm free})}(x,x'\vert s)$ are given in Appendix
{\ref{A:free-green-function}}.

We also wish to point out that for static and homogeneous background
fields ($v_0$) the computation of $G_\field[v_0]$ and $G_\ghost[v_0]$
is not much more complicated than that for $G_\free$. Details are
presented in Appendix {\ref{A:homogeneous}}. These static and
homogeneous calculations allow one to compare with the effective
potential formalism of~\cite{priority2c}, and serve as a check on the
current effective action calculation.

We adopt the following ansatz to perturbatively solve the heat
equations (\ref{E:schrodinger2}) for small $s$ (adiabatic
approximation)
\begin{mathletters}
\begin{eqnarray}
\label{E:ansatz-field}
G_\field(x,x'|s) &=& 
G_\free(x,x'|s) \; f_\field(x,x'|s), 
\\
\label{E:ansatz-ghost}
G_\ghost(x,x'|s)&=& 
G_\free(x,x'|s) \; f_\ghost(x,x'|s),
\end{eqnarray}
\end{mathletters}
where
\begin{mathletters}
\begin{eqnarray}
\label{E:f-field}
f_\field(x,x'|s) = \sum_{l=0}^{+\infty} s^{\frac{l}{2}} \; 
b_{\frac{l}{2}}(x,x') 
= b_0 + s^{\half}b_{\half} 
+ s \; b_1 + O(s^{\frac{3}{2}}),
\\
\label{E:f-ghost}
f_\ghost(x,x'|s) = \sum_{l=0}^{+\infty} s^{\frac{l}{2}} \; 
a_{\frac{l}{2}}(x,x')  
= a_0 + s^{\half}a_{\half} 
+ s \; a_1 + O(s^{\frac{3}{2}}),
\end{eqnarray}
\end{mathletters}
are asymptotic series in half-integer powers of the proper time with
coefficient functions $b_{\frac{l}{2}}$ and $a_{\frac{l}{2}}$ (called
``Seeley--DeWitt'' coefficients).  Note that we have had to consider
fractional powers in this small $s$ expansion. [By considering simple
cases it is easy to convince oneself that for a differential operator
of order $n$ the ``heat kernel'' expansion should start with an
overall factor proportional to $s^{-d/n}$ and then contain subdominant
terms that are integer powers of $s^{2/n}$.]

{\emph{In principle}}, these coefficients can be calculated to
arbitrarily high order by solving recursion relations obtained by
substituting (\ref{E:ansatz-field}) and (\ref{E:ansatz-ghost}) into
(\ref{E:schrodinger2}) for $H_\field$ and $H_\ghost$, respectively.
The boundary (initial) conditions $G_\field(x,x'|0)= G_\ghost(x,x'|0)=
\delta^n(x,x')$ imply that $b_0 = 1$ and $a_0=1$. These coefficients
start the Seeley--DeWitt hierarchy and allow for a complete
determination of the Seeley--DeWitt coefficients appearing in
(\ref{E:f-field})--(\ref{E:f-ghost}).  For second-order differential
operators this procedure has now become
automated~\cite{Schimming}. For fourth-order differential operators
considerably less is known~\cite{Gusynin}.

{\emph{In practice}}, we will see that only the first Seeley--DeWitt
coefficients are germane to the problem and that it is sufficient to
find the ``integrated'' Seeley--DeWitt coefficients. This permits us
to calculate the relevant coefficients by means of a technique based
on the Feynman--Hellman formula~\cite{Feynman-Hellman}, which can
itself be viewed as a specialization of the Baker--Campbell--Hausdorff
formula~\cite{BCH}.

We regulate the one-loop effective action by cutting off the lower
limit of the proper time integral
\begin{equation}
\label{E:regulated}
\Gamma_{\epsilon}^{(1-loop)}[\phi] \equiv
-\frac\A{2}\int_{\epsilon}^{+\infty}\frac{\d s}{s} \; 
\Tr\left(e^{-sH_\field} - e^{-s{[H_0]}_\field} \right)
+
\frac\A{2}\int_{\epsilon}^{+\infty}\frac{\d s}{s} \; 
\Tr\left(e^{-sH_\ghost} - e^{-s{[H_0]}_\ghost} \right),
\end{equation}
where we can identify $\epsilon = 1/{\Omega^2_\cut-off}$ and
$\Gamma_{\epsilon}^{(1-loop)}[\phi]$ with
$\Gamma_{\Omega_\cut-off}^{(1-loop)}[\phi]$. As the product $sH$ must
be dimensionless, we deduce that $s$ has engineering dimensions of
(time)$^2$ or equivalently, (frequency)$^{-2}$.  In this stochastic
field theory the cut-off can be taken to be a frequency scale
$\Omega_\cut-off$, and this identification allows us to compare
between these two types of cut-off (proper time versus
frequency). Since these theories are {\emph{not}} Lorentz invariant, a
frequency cut-off is not ``quite'' interchangeable with a wavenumber
cut-off (more on this point below).

Substituting the above ansatz
(\ref{E:ansatz-field})--(\ref{E:ansatz-ghost}) into
(\ref{E:regulated}), making use of the explicit form for $G_\free$
(\ref{E:formal}), and expanding out the first terms of $f_\field$ and
$f_\ghost$ yields
\begin{eqnarray}
\frac{\Gamma_{\epsilon}^{(1-loop)}[\phi]}{
\cal A} 
&=&
-\half
\int_{\epsilon}^{+\infty}
\frac{\d s}{s}\; 
\int \d^n{x}\;
[G_\free(x,x'|s)] 
\; 
\left( [f_\field(x,x'|s)]  - [f_\ghost(x,x'|s)]
\right)
\nonumber \\
&=& 
-\half
\int_{\epsilon}^{+\infty}
\frac{\d s}{s} \; 
\int \d^n{x}\; 
[G_\free(x,x'|s)]  
\nonumber \\
&&
\qquad \qquad 
\left(
s^{\half}\;[b_{\half} (x,x') ] -
s^{\half}\;[a_{\half} (x,x')  ]
+ s \;[b_1 (x,x')]  - s \; [a_1 (x,x')] 
+ O(s^{\frac{3}{2}}) \right)
\nonumber \\
&=& 
- \frac{1}{2}
A_d
\int_{\epsilon}^{+\infty}
\frac{\d s}{s^{\frac{3}{2} + \frac{d}{4}} } \;
\int \d^n  x\; 
\left(
s^{\half}\;[c_{\half}(x,x')] 
+ s \; [c_1 (x,x')] + O(s^{\frac{3}{2}})\right).
\end{eqnarray}
In arriving at this last expression we have used the fact that the
coincidence limit of the free heat kernel is
\begin{equation}
[G_\free(x,x'|s)] = 
A_d \;
s^{-\half - \frac{d}{4}},
\end{equation}
which follows immediately from (\ref{E:formal}).  We have also made
use of the standard notation to express coincidence limits.  Given any
function $h(x,x')$, we write
\begin{equation}
\lim_{x' \rightarrow x} h(x,x') = [h(x,x')].
\end{equation}
(Although we also employ the square brackets to denote arguments of
functionals and functions, and to group expressions, the intended
meaning should be clear from context and there should be no
confusion.)  We have defined {\emph{net}} Seeley--DeWitt coefficients
\begin{equation}
c_{\frac{l}{2}} \equiv b_{\frac{l}{2}} - a_{\frac{l}{2}} , \; \; \; \forall \; 
l=0,1,2,\dots .
\end{equation}
The Seeley--DeWitt coefficients $b_{\frac{l}{2}}$, $a_{\frac{l}{2}}$,
and $c_{\frac{l}{2}}$ are functions of the mean field $\phi(\vec x,t)$
and its derivatives, and as remarked above, can in principle be
determined by solving a recursion relation resulting from inserting
the ansatz (\ref{E:ansatz-field}) and (\ref{E:ansatz-ghost}) into
(\ref{E:schrodinger2}).  However, to obtain the {\emph{form}} of the
divergences of the one-loop effective action we need not evaluate
these coefficients.  It suffices to calculate the lower bound $(s
\rightarrow 0)$ of the proper time integral.  In the limit $\epsilon
\rightarrow 0$ we find that the divergent terms in the RD effective
action are given by
\begin{eqnarray}
\label{E:divergences}
\Gamma_{\epsilon}^{(1-loop)}[\phi] &=& 
- \frac{1}{2}A_d {\cal A}
\left( \frac{4}{d}\; \epsilon^{-\frac{d}{4}} \int \d^n x\;
[c_{\half}(x,x')]
-
\frac{\epsilon^{\half -\frac{d}{4}}}{(\half - \frac{d}{4})}
\int \d^n x \; [c_1(x,x')] 
\right. 
\nonumber \\
&-& 
\left.
\frac{\epsilon^{1-\frac{d}{4}}}{(1- \frac{d}{4})}
\int \d^n x \; [c_{\frac{3}{2}}(x,x')]
 + \; \cdots \right).
\end{eqnarray}

We now list the divergences in the RD one-loop effective action for
the following space dimensions
\begin{mathletters}
\begin{eqnarray}
&\hbox{\underline{$d=0$}}& 
\qquad\qquad\qquad
\label{E:zero-dim}
\Gamma_{\epsilon}^{(1-loop)}[\phi] 
= - 2 A_0 \; {\cal A}
\; \log(\Omega^2 \; \epsilon) 
\int \d t\; [c_{\half}],
\\
&\hbox{\underline{$d=1$}}&
\qquad\qquad\qquad
\label{E:one-dim}
\Gamma_{\epsilon}^{(1-loop)}[\phi] 
=
- 2 A_1 \; {\cal A}
\; \epsilon^{-\frac{1}{4}}
\int \d x \; \d t\; [c_{\half}],
\\
&\hbox{\underline{$d=2$}}&
\qquad\qquad\qquad
\label{E:two-dim}
\Gamma_{\epsilon}^{(1-loop)}[\phi] 
=
-\frac{A_2}{2} \; {\cal A}
\left(2 \epsilon^{-\half} \int \d^2\vec x \; \d t\; [c_{\half}] 
-\log(\Omega^2 \; \epsilon) 
\int \d^2\vec x \; \d t\; [ c_1] \right),
\\
&\hbox{\underline{$d=3$}}&
\qquad\qquad\qquad
\label{E:three-dim}
\Gamma_{\epsilon}^{(1-loop)}[\phi] 
=
-\frac{A_3}{2} \; {\cal A}
\left( \frac{4}{3}\epsilon^{-\frac{3}{4}} 
\int \d^3\vec x \; \d t\; [ c_{\half} ]
+ 4 \epsilon^{-\frac{1}{4}}
\int \d^3\vec x \; \d t\; [c_1] \right).
\end{eqnarray}
\end{mathletters}
In all of these cases we only need to solve for the first two
adiabatic (Seeley--DeWitt) coefficients $c_{\half}$, and $c_1$;
indeed, it is only the spacetime integrated {\emph{net}} coefficients
that are needed.  In higher {\emph{space}} dimensions,
{\emph{additional}} $c_{\frac{l}{2}}$'s would be required. However,
for most practical applications it is enough to consider $0 \leq d\leq
3$. (Moreover, earlier work regarding the one-loop effective potential
for RD indicates that this field theory is non-renormalizable for $d
\geq 4$~\cite{priority2c}).  This dimension range covers the spatially
homogeneous limit ($d=0$), one-dimensional (linear) systems ($d=1$),
surfaces ($d=2$), and bulk systems (volumes) ($d=3$).  We see that the
divergences are of two types: (fractional) powers of the cut-off and
logarithms of the cut-off. Only the latter can yield one-loop
renormalization group beta functions and associated RGEs.

\section{Calculation of the Seeley--DeWitt coefficients}

In this section we present a formalism that can {\emph{in principle}}
yield all the Seeley--DeWitt coefficients.  As we have seen in the
previous section, the calculation of the one-loop effective action
involves solving the auxiliary partial differential equations of
parabolic type (denoted as heat equations, even if the diffusion is
non-standard).

In this formalism (see Appendix \ref{A:feynman-hellman}) the first
step is to write
\begin{equation}
\Tr[\exp(-s H)] = \Tr\{\exp[-s(H_\free + \delta H)]\},
\end{equation}
where $\delta H$ is a lower-order differential operator when compared
to $H$ or $H_\free$. We now apply a version of the Feynman--Hellman
formula~\cite{Feynman-Hellman,BCH}, as discussed in Appendix
{\ref{A:feynman-hellman}}, to obtain
\begin{eqnarray}
\Tr [ \exp (A+\e B)]
&=& 
\Tr[\exp (A)] + 
\e \Tr [ B \; \exp(A)] 
+
{\e^2\over2} \int_0^1 \d\ell\;
\Tr\left\{ B \; \exp(\ell A)  \; B  \; \exp[(1-\ell)A])\right\} 
+
O(\e^3).
\label{E:f-h1}
\end{eqnarray}
This equation will be the basis for extracting the first two
{\emph{integrated}} Seeley--DeWitt coefficients.

The second step is to realise that we only need to look at the
difference
\begin{equation}
\Tr[\exp(-s H_\field)] - \Tr[\exp(-s H_\ghost)],
\end{equation}
and write
\begin{mathletters}
\begin{eqnarray}
\Tr[\exp(-s H_\ghost)] &=& \Tr\{\exp[-s(H_\free + \delta H_1)]\},
\\
\Tr[\exp(-s H_\field)] &=& \Tr\{\exp[-s(H_\free + \delta H_1 + \delta H_2)]\}.
\end{eqnarray}
\end{mathletters}
If we take the difference of the previous operators, the $O(\e^0)$
term automatically cancels, as does the $O[\e^2 (\delta H_1)^2]$ term,
leaving
\begin{eqnarray}
\Tr[\exp(-s H_\field)] - \Tr[\exp(-s H_\ghost)]
&=&  
- \e s \Tr [ \delta H_2 \; \exp(-s H_\free)] 
\nonumber\\
&& +
{\e^2\over2} s^2 \int_0^1 \d\ell \;
\Tr\left\{ 
\delta H_2 \; \exp(-\ell sH_\free)  \; \delta H_2  \; \exp[(1-\ell)sH_\free])
\right\} 
\nonumber\\
&& +
\e^2 s^2 \int_0^1 \d\ell \;
\Tr\left\{ 
\delta H_1 \; \exp (-\ell sH_\free)  \; \delta H_2  \; \exp[(1-\ell)sH_\free])
\right\} 
\nonumber\\
&& +
O(\e^3).
\end{eqnarray}
We now make use of the explicit form of these ``Hamiltonians''. We write
\begin{eqnarray}
H_\ghost &=& 
(D^\dagger-V')(D-V')
=
D^\dagger D -  (D^\dagger-V')  V'  + 2 \nu \nabla V' \cdot \nabla,
\end{eqnarray}
where in the last term of the right hand side both $\nabla$'s act on
{\emph{everything}} to the right. We know that the free Hamiltonian is
given by
\begin{equation}
H_\free = D^\dagger D = [-\partial_t^2 + \nu^2 (\nabla^2)^2],
\end{equation}
so that we can identify $\delta H_1$ as
\begin{equation}
\delta H_1 
= - [ (D^\dagger-V')  V' - 2 \nu \nabla V' \cdot \nabla ].
\end{equation}
(Note that $\delta H_1$ is a linear differential operator.)  {From}
the definition of the ghost Hamiltonian
\begin{equation}
H_\field = H_\ghost - V'' (D\phi - V),
\end{equation}
we deduce the following form for $\delta H_2$
\begin{equation}
\delta H_2 = - V'' (D\phi - V).
\end{equation}
(Note that $\delta H_2$ is a function, not a differential operator.)
Now consider the first order perturbation [the $O(\e)$ term]
\begin{equation}
X_1 \equiv  s \; \Tr\{
\left[ V''(D\phi-V)  \right]
\; \exp(-sH_\free)
\}.
\end{equation}
{From} the known form of the free kernel, [see, \eg, equation
(\ref{E:formal}) or equation (\ref{E:master})], and the fact that
$\delta H_2$ is a function, this reduces to
\begin{equation}
X_1 = A_d \; s^{-{1\over2}-{d\over4}} \int  \d^n x   \;
\left[ s \;  V'' (D\phi - V)   \right].
\end{equation}
This implies that the first-order perturbation does not contribute to
the Seeley--DeWitt coefficient $c_{\frac{1}{2}}$, though it does
contribute to $c_1$. In fact, we can write
\begin{equation}
\int \d^n x   \; [c_{1}] = 
\int \d^n x   \; V'' (D\phi-V) + \cdots.
\end{equation}
This is actually the {\emph{only}} contribution to the relevant
Seeley--DeWitt coefficients.  (There might have been additional
contributions coming from those portions of the second-order term
$X_2$ that have space derivatives; fortunately they vanish, as we now
verify.)  Let us consider
\begin{eqnarray}
X_2
&\equiv&  
 {s^2\over2}   \int_0^1 \d\ell \; \Tr\big\{
\left[ V'' (D\phi - V)  \right] 
\; \exp (-s\;\ell\;H_\free)
\left[  V'' (D\phi - V) \right] 
\; \exp[-s\;(1-\ell)\;H_\free] \;\big\}
\\
&+&
 s^2   \int_0^1 \d\ell \; \Tr\big\{
\left[  V'' (D\phi - V) \right] 
\; \exp(-s\;\ell\;H_\free)
\left[ 
(D^\dagger-V')  V' - 2 \nu (V' \nabla^2 + (\nabla V')\cdot \nabla ) 
\right] 
\; \exp[-s\;(1-\ell)\;H_\free] \;\big\}.
\nonumber
\end{eqnarray}
We can have any of the following cases

-- No gradients hit the free kernel: the term containing two factors
of $[V'' (D\phi - V)]$ is proportional to $s^2$ and can only
contribute to $c_2$, which is not needed in the present context.

-- One gradient hits one kernel: from equation (\ref{E:formal}) or
equation (\ref{E:master}), one can see that there is a factor of
$(x-x')_i/(\nu\sqrt{s})$ of order $s^{3/2}$. Such a term is odd under
the interchange of $x$ and $x'$ and will vanish when taking the
spacetime trace.

-- Two gradients hit the same free kernel: there will be contributions
of the type
\begin{equation}
C_{d,1} {\delta_{ij}\over\nu\sqrt{s}} + 
(C_{d,1})^2 {(x-x')_i \; (x-x')_j\over d^2 s},
\end{equation}
which, after tracing with the free kernel, yield contributions
proportional to $s^{3/2}$. Therefore, these terms contribute to
$c_{\frac{3}{2}}$, which is not needed.

Continuing in this manner, it is easy to convince oneself that there
are no additional contributions to the required coefficients
$c_{\frac{1}{2}}$ and $c_{1}$. We can finally write
\begin{mathletters}
\begin{eqnarray}
\label{E:c-half}
&&\int \d^n x \; [c_{\frac{1}{2}}] = 0,
\\
\label{E:c1}
&&\int \d^n x \; [c_{1}] = \int \d^n x  \;
 [ V'' (D\phi-V) ].
\end{eqnarray}
\end{mathletters}
Note that the present calculation only yields the {\emph{integrated
on-diagonal}} ($x=x'$) Seeley--DeWitt coefficients and is insensitive
to any term that vanishes upon integration.

With a little more work along these lines, it is also possible to
obtain the Seeley--DeWitt coefficients: $[a_{\frac{1}{2}}]$ and
$[a_{1}]$ for the ghost field, and $[b_{\frac{1}{2}}]$ and $[b_{1}]$
for the physical field. We only quote the results here
\begin{mathletters}
\begin{eqnarray}
\label{E:a-half}
&&
\int \d^n x  \; [a_{\frac{1}{2}}] = \int \d^n x  \;   2 \; C_{d,1} \; V',
\\
\label{E:a1}
&&
\int \d^n x  \;  [a_1] = 
\int \d^n x  \;  
\left[ 
{d\over 2}\;\left( V' \right)^2   + (D^\dagger-V')V'
\right],
\\
\label{E:b-half}
&&
\int \d^n x  \;  [b_{\frac{1}{2}}] = \int \d^n x  \;  2 \; C_{d,1} \; V',
\\
\label{E:b1}
&&
\int \d^n x  \;  [b_1] = \int \d^n x  \; 
\left[
{d\over 2}\;\left( V' \right)^2   
+ (D^\dagger-V')V' + V'' (D\phi-V)
\right].
\end{eqnarray}
\end{mathletters}
%

\section{One-loop Renormalization}

We have already calculated the (regularized) one-loop effective action
(\ref{E:effective2}) for the RD equation, and thus, we may now explore
the renormalizability of this field theory following the prescription
reviewed in~\cite{phi4}.  In order to do so we must analyze the
divergences of the one-loop effective action
$\Gamma_{\epsilon}^{(1-loop)}$. We must also keep in mind the fact
that the bare theory (\ie, defined by the action (\ref{E:action3}))
does not depend on the arbitrary scale $\mu$ introduced by the
renormalization scheme. Therefore, just as for the case of
QFTs~\cite{Fujimoto,Gato}, we will derive a set of equations that
govern the scale dependence of the parameters appearing in the RD
effective action from the identity
\begin{equation}
\label{E:RGidentity}
\mu \frac{\d \Gamma[\phi]}{\d \mu} = 0 =
\mu \frac{\d (S[\phi] - S[v_0]) }{\d \mu} +
 \mu  \frac{\d \Gamma_{\epsilon}^{(1-loop)}[\phi]}{\d \mu} 
+ O(\A^2),
\end{equation}
where the $O(\A^2)$ indicates there will be higher-loop contributions
to the effective action. In quantum field theory this identity does
yield the one-loop RGEs since equation (\ref{E:RGidentity}) can be
expressed in terms of a sum of independent field operators (operator
basis) and each coefficient of an element of this basis determines an
independent RGE.

As we have already calculated the relevant Seeley--DeWitt coefficients
for the RD equation, we now turn to investigate the one-loop
renormalizability of its stochastic field theory.  The
renormalizability criteria are based on the following definitions. For
renormalizable and super-renormalizable theories the counterterms
needed to cancel the divergences are equal to, or fewer in number than
the terms appearing in the zero-loop action, which implies that the
Seeley--DeWitt coefficients are expandable in terms of the
{}\emph{same} operator basis appearing in the classical action.  In
particular, this basis set consists of $\{ \partial_t \phi, \nabla^2
\phi, 1,\phi,\phi^2, \cdots, \phi^N \}$.  For non-renormalizable
theories this criterion fails. That is, there are terms in the
integrated Seeley--DeWitt coefficients that do not appear in the
classical action~\cite{Ramond,Zinn,BVW-zeta,priority1a}.

By comparing the zero-loop action (\ref{E:action3}) with the divergent
terms of the one-loop effective action, it is easy to see that the
latter do not contain any field operators not already present in the
bare (classical) action.  The divergent contributions to the one-loop
effective action in $d$-dimensions are given by
\begin{eqnarray}
\label{E:netaction}
\Gamma^{(1-loop)}_{\epsilon}[\phi;v_0]
&=&\frac\A{2} \; A_d \; 
\frac{\epsilon^{\half-\frac{d}{4}}}{(\half-\frac{d}{4})}\; 
\int \d^dx \; \d t\; 
[c_1(\phi)-c_1(v_0)]
+ O(\epsilon^{1-\frac{d}{4}})
\nonumber \\
&=&\frac\A{2} \; A_d \; 
\frac{\epsilon^{\half-\frac{d}{4}}}{(\half-\frac{d}{4})}\; 
\int \d^dx \; \d t\; 
\{V''(\phi)[D\phi-V(\phi)] +V''(v_0) V(v_0)\}
+ O(\epsilon^{1-\frac{d}{4}}).
\end{eqnarray}
Some remarks are in order. First of all we point out the fact that the
one-half Seeley--DeWitt coefficients of the field and ghost mutually
cancel out.  This cancellation is special to the RD system and does
not take place for generic SPDEs.  Secondly, the ill-defined quantity
``$\epsilon^0/0$'' arising in $d=2$ must be replaced by
$\log(\Omega^2\epsilon) = \log(\Omega^2/\Omega_\cut-off^2)$. The
dimensionfull (but arbitrary) parameter $\Omega$ is required to make
the argument of the logarithm dimensionless. It is often more
convenient to introduce a cut-off in wavenumber, rather than in
frequency. In Lorentz invariant theories (QFTs, for example) these are
essentially equivalent and it is usual to adopt units where the speed
of light is one. In the RD system this would be inappropriate, since
the equation is not Lorentz invariant.  Instead, one notes that
dimensionally
\begin{equation}
[\epsilon] = [\hbox{proper time $s$}] = 
[\hbox{physical time $t$}]^2 = 
[\nu]^{-2} [\hbox{distance}]^4,
\end{equation}
and therefore, in terms of a wavenumber cut-off $\Lambda$ and a
wavenumber subtraction point $\mu$
\begin{equation}
\hbox{``$\epsilon^0/0$''} \to  
\log(\Omega^2\epsilon) =
\log(\Omega^2/\Omega_\cut-off^2) =  
\log(\mu^4/\Lambda^4).
\end{equation}
(This observation is important when comparing different regularization
schemes; for instance the effective potential calculation
of~\cite{priority2c} uses a wavenumber cut-off.)

Thirdly, it is of great importance to study the {}\emph{type} of
divergence arising in the one-loop effective action for the RD
equation, \ie, logarithms versus (fractional) power.  {From} the above
we see that the type of divergence depends on the number of
{}\emph{space} dimensions $d$.  If $d$ is odd, there will never be
logarithmic divergences to one-loop order in the RD field theory; to
get a logarithm, the space dimensionality must be even.  A similar
feature holds also for the one-loop effective action for QFTs in odd
{}\emph{spacetime} dimensions~\cite{phi4,priority1a}.  We can conclude
that the appearance of logarithmic divergences for specific space
dimensions is not an artifact of the RD field theory, nor of SPDEs,
nor is it an artifact of the regularization scheme we have employed.
In QFT the RGEs yield the running of the coupling constants, \ie, give
the scale dependence, if and only if, there are logarithmic
divergences in the effective action. Thus, we can already predict that
the parameters in the RD equation will not run in the ultraviolet
region (to one-loop order) for {\emph{odd}} space
dimensions~\cite{priority1a}.

Nevertheless, the one-loop effective action in odd space dimensions
still contains divergences (though not for $d=1$), which must be
subtracted by suitable counterterms, but once this subtraction has
been performed, the remaining finite part $\Gamma^{(1-loop)}_{finite}$
is independent of the subtraction scale $\mu$.

\section{Renormalization of the RD effective action}

In this section we calculate the counterterms needed to renormalize
the one-loop effective action.  We first start with the bare classical
action (\ref{E:action3}) for the reaction-diffusion equation
\begin{eqnarray}
S[\phi]&=& \frac{1}{2} \int \d^n x \left( 
\partial_t \phi 
- \nu \nabla^2 \phi
-V[\phi]
\right)^2
=
 \frac{1}{2} \int \d^n x \left( 
\partial_t \phi 
- \nu \nabla^2 \phi
- \sum_{j=0}^{N} \frac{\lambda_j}{j!} \phi^j
\right)^2.
\end{eqnarray}
The bare parameters (no subscript) are related to the renormalized
ones (denoted by a subscript $R$) as follows
\begin{eqnarray}
\phi&=& Z^{-1/2}\phi_R
, \; \; \; {\rm with} \; \; \; Z = 1 + \delta Z ,
\\
\lambda_j&=& \lambda^R_j + \delta \lambda_j
,
\\
\nu &=& \nu_R + \delta \nu
,
\end{eqnarray}
with $\delta Z, \delta \lambda_j$, and $\delta \nu$ the corresponding
counterterms for $Z,\lambda_j$, and $\nu$, respectively. (Our
convention for the definition of the wavefunction renormalization
constant $Z$ does not follow the standard one in
QFT~\cite{Ramond,Zinn}.)

Our task consists in demonstrating that all the divergences appearing
in the regulated one-loop effective action can be cancelled by
suitable choices for these counterterms. In effect, we absorb the
divergences into the (bare) parameters of the RD equation by
{}\emph{renormalizing} these parameters.  If we write the bare action
in terms of the renormalized parameters and counterterms, and keep up
to linear order in the counterterms, (which is sufficient for a
one-loop analysis), we find
\begin{eqnarray}
S[\phi]&=& \frac{1}{2} 
\int \d^n x \left( 
\partial_t \phi_R 
- \nu_R \nabla^2 \phi_R
- \sum_{j=0}^{N} \frac{\lambda_j^R}{j!} \phi^j_R 
\right)^2
- \frac{\delta Z}{2}
\int \d^n x \left( 
\partial_t \phi_R 
- \nu_R \nabla^2 \phi_R
-  \sum_{j=0}^{N} \frac{\lambda_j^R}{j!} \phi^j_R  
\right)^2
\nonumber \\
&-& \int \d^n x \left( 
\partial_t \phi_R 
- \nu_R \nabla^2 \phi_R
- \sum_{j=0}^{N} \frac{\lambda_j^R}{j!} \phi^j_R 
\right) 
\left(
\delta \nu \, \nabla^2 \phi_R
+ \sum_{j=0}^{N}   \frac{ \phi^j_R}{j!}
\left[\delta \lambda_j + \delta Z \frac{\lambda_j^R(1-j)}{2}  \right]
\right),
\nonumber
\end{eqnarray}
which can be written in a more compact and transparent notation as
follows
\begin{eqnarray}
S[\phi]&=& \frac{1}{2} 
\int \d^n x \left( 
D_R \phi_R - V_R[\phi_R]
\right)^2
- \frac{\delta Z}{2}
\int \d^n x \left( 
D_R \phi_R - V_R[\phi_R]
\right)^2
\nonumber \\
&-&
 \int \d^n x \left( 
D_R \phi_R - V_R[\phi_R]
\right)
\left(
\delta \nu \, \nabla^2 \phi_R
+ \sum_{j=0}^{N}   \frac{ \phi^j_R}{j!}
\left[\delta \lambda_j + \delta Z \frac{\lambda_j^R(1-j)}{2}  \right]
\right)
\label{E:bare-action}
\\
\label{E:ren-delta-action}
&=& S_R[\phi_R,\mu] + S_{\delta}[\phi_R],
\end{eqnarray}
where we have introduced the scale dependent renormalized action
$S_R[\phi_R,\mu]$
\begin{eqnarray}
S_R[\phi_R,\mu] &\equiv& \half\int \d^n x\; \Big(
Z^{\half}(\mu) \partial_t \phi
-\nu_R (\mu) Z^{\half}(\mu) \nabla^2 \phi
-V[Z^{\half}\phi,\lambda_j^R(\mu)] \Big)^2,
\end{eqnarray}
with 
\begin{equation}
V_R[\phi_R] \equiv 
V[Z^{\half}\phi,\lambda_j^R(\mu)] = \sum_{j=0}^N 
\frac{\lambda_j^R(\mu)}{j!}Z^{\frac{j}{2}}(\mu) \phi^j,
\end{equation}
the renormalized (scale dependent) potential.

The meaning of $D_R$ is clear from inspection. The final equality in
(\ref{E:ren-delta-action}) defines the finite, renormalized action
$S_R$ and the divergent (but regulated) counterterm action
$S_{\delta}$.  The individual counterterms appearing in $S_{\delta}$
will be used to cancel off the divergences arising in
(\ref{E:netaction}).  We carry out this cancellation separately in
each space dimension since each case leads to structurally different
divergences [see equation (\ref{E:netaction})].

\subsection{$d=0$ counterterms and renormalization}

The case $d=0$ is very simple: in zero space dimensions there is no
diffusion. There is a brief discussion in reference~\cite{priority2c}
and we do not belabor the point here except to reiterate that in $d=0$
the RD equation is one-loop renormalizable and one-loop finite.

\subsection{$d=1$ counterterms and renormalization}

In one space dimension the (formally) divergent effective action is
given by
\begin{eqnarray}
\Gamma_\epsilon^{(1-loop)}[\phi_R]&=& - 2 \A A_1 \epsilon^{-1/4}
\int \d x \; \d t \; [c_{\half}] + O(\A^2),
\end{eqnarray}
with $A_1={\Gamma(1/4)}/[{8 \pi (\pi \nu_R)^{1/2}}]$.  We did not
explicitly write this term in equation (\ref{E:netaction}) since it
vanishes identically.

{From} our previous calculation of the Seeley--DeWitt coefficients we
know that $[c_{\half}]= 0$ (in all dimensions), which tells us that in
one space dimension there are no divergences, that is, the theory is
one-loop finite and there is no need to introduce counterterms.  Since
no renormalization is required, there will be no scale dependence in
the parameters appearing in the RD equation.  The beta functions,
$\beta_{\cal O} \equiv \mu\, d{\cal O}/d\mu$, (encoding the scale
dependence of the parameters) are therefore zero [at least up to order
$O(\A^2)$], and we have
\begin{eqnarray}
Z &=& 1 + O(\A^2),
\\
\nu &=& \nu_R + O(\A^2), 
\\
\lambda_j &=& \lambda^R_j + O(\A^2).
\end{eqnarray}

\subsection{$d=2$ counterterms and renormalization}

In two space dimensions the divergent effective action is given by
equation (\ref{E:netaction})
\begin{eqnarray}
\Gamma_\epsilon[\phi_R]&=& +\frac{\A A_2}{2} \log (\mu^4/\Lambda^4)
\int \d^2 \vec x \; \d t \; [c_{1}],
\end{eqnarray}
where $A_2=1/({16 \pi \nu_R})$. 

{From} the calculation of the Seeley--DeWitt coefficients we know that
for $d=2$ we have
\begin{mathletters}
\begin{eqnarray}
\label{E:c-half-2}
[c_{\half}]&=& 0,
\\
\label{E:c-one-2}
[c_1]
&=&
V_R''[\phi_R](D\phi_R - V_R[{\phi_R}]),
\end{eqnarray}
\end{mathletters}
where we have written the Seeley--DeWitt coefficient $[c_1]$ in terms
of the renormalized parameters as we are only working to one-loop
order.  Therefore, for the divergent part of effective action we can
write
\begin{eqnarray}
\Gamma_\epsilon^{(1-loop)}[\phi_R]&=&
\frac{\A}{8\pi\nu_R} \log (\mu/\Lambda)
\int \d^2 \vec x \; \d t \; 
V_R''[\phi_R](D_R \phi_R - V_R[\phi_R]).
\end{eqnarray}
In order to determine the value of the counterterms 
and to cancel them off, we must set
\begin{eqnarray}
\Gamma_\epsilon^{(1-loop)}[\phi_R]&=& - S_{\delta}[\phi_R]
+ {\rm finite}.
\end{eqnarray}
This cancellation can be made up to a residual finite but scale
dependent logarithm. This is because the difference of two divergent
logarithms can be finite and {}\emph{non-zero}. The counterterms are
proportional to $\log(\mu / \Lambda)$, where $\mu$ is an arbitrary
scale needed to render the argument dimensionless, but this scale need
not coincide with $\mu_0$, the {}\emph{other} arbitrary scale needed
to render the argument of the other logarithm, $\log(\mu_0 /\Lambda)$,
dimensionless~\cite{priority1a}.

If we perform this cancellation, we obtain the following
$\mu$-dependent family of solutions for the counterterms $\delta
\lambda_j$
\begin{eqnarray}
\frac{\A}{2}A_2\log (\mu^2 /\Lambda^2) \; V''_R[\phi] =
\frac{\A}{2}A_2\log (\mu^2 /\Lambda^2) \;
\sum_{j=0}^{N}\frac{\lambda_j^R}{j!}j(j-1)\;
\phi^{j-2}&=& \sum_{j=0}^{N}\frac{\delta \lambda_j}{j!} \; \phi^j.
\end{eqnarray}
As we are working to one-loop order, we can set $Z(\mu)$ equal to one
in $V''_R$.  We can then read off the individual counterterms from
this equation, using the linear independence of the basis $\{
\partial_t \phi, \nabla^2 \phi, 1,\phi,\phi^2, \cdots, \phi^N \}$, to
obtain
\begin{mathletters}
\begin{eqnarray}
\delta Z &=& 0 +O(\A^2),
\\
\delta \nu &=& 0  +O(\A^2), 
\\
\label{E:explicit-2d}
\delta \lambda_0 &=& 
\frac{\A}{8\pi\nu_R} \log (\mu/\Lambda) \lambda^R_2  +O(\A^2), 
\\
\delta \lambda_1 &=& 
\frac{\A}{8\pi\nu_R} \log (\mu/\Lambda) \lambda^R_3 +O(\A^2), 
\\
\vdots
\nonumber
\\
\delta \lambda_{N-2} &=& 
\frac{\A}{8\pi\nu_R} \log (\mu/\Lambda) \lambda^R_{N}  +O(\A^2), 
\\
\delta \lambda_{N-1} &=& 0 +O(\A^2), 
\\
\delta \lambda_{N} &=& 0 +O(\A^2) .
\end{eqnarray}
\end{mathletters}
In particular, we see that there is no wavefunction renormalization
nor diffusion constant renormalization in two dimensions at one-loop
order.  The couplings associated with the highest and next-to-highest
powers of the field $(\lambda^R_{N-1},\lambda^R_N)$ do not require
renormalization to this order.

As pointed out above, due to the logarithmic divergence in two
dimensions, when we subtract the divergences from the counterterm
action, we are left with a finite $\mu$-dependent logarithmic piece in
addition to the renormalized action, that is
\begin{eqnarray}
\Gamma[\phi] &=& S[\phi] + \Gamma_{\epsilon}^{(1-loop)}[\phi] +
\Gamma_{finite}^{(1-loop)}[\phi] \nonumber 
\\ 
&=& S_R[\phi_R,\mu] +
S_{\delta}[\phi_R] + \Gamma_{\epsilon}^{(1-loop)}[\phi_R] +
\Gamma_{finite}^{(1-loop)}[\phi_R] \nonumber 
\\ 
&=& S_R[\phi_R,\mu] +
{\A \, A_2 \over 2} \log \left(\frac{\mu_0^4}{\mu^4} \right) \int \d^2
\vec x \; \d t \; (D_R \phi_R - V_R[\phi_R]) V''_R[\phi_R] +O(\A^2).
\end{eqnarray}
We insert this expression into (\ref{E:RGidentity}) to obtain the
one-loop RGE
\begin{equation}
\label{E:bigRGE}
\mu \frac{\d}{\d \mu} (D_R\phi_R -V_R[\phi_R]) = {\A \over 8\pi \nu_R}
V''_R[\phi_R]  +O(\A^2).
\end{equation}
In arriving at this equation, we have cancelled an overall
{}\emph{common} factor of the classical equation of motion, since
$D_R\phi_R - V_R[\phi_R] \neq 0$ in general.

By collecting up the coefficients of the linearly independent terms in
(\ref{E:bigRGE}) we find the corresponding one-loop RGEs and beta
functions in $d=2$ to be given by
\begin{mathletters}
\begin{eqnarray}
\beta_Z &=& \mu \frac{\d Z}{\d \mu} 
= 0  +O(\A^2),
\\
\beta_\nu &=& \mu \frac{\d \nu_R}{\d \mu} 
= 0  +O(\A^2), 
\\
\beta_{\lambda_0} &=& \mu \frac{\d \lambda^R_0}{\d \mu} 
=- {\A \over 8\pi \nu_R} \lambda^R_{2}  +O(\A^2), 
\\
\beta_{\lambda_1} &=& \mu \frac{\d \lambda^R_1}{\d \mu} 
=- {\A \over 8\pi \nu_R} \lambda^R_{3}  +O(\A^2), 
\\
\vdots
\nonumber
\\
\beta_{\lambda_{N-2}} &=& \mu \frac{\d \lambda^R_{N-2}}{\d \mu} 
=- {\A \over 8\pi \nu_R} \lambda^R_{N}  +O(\A^2), 
\\
\beta_{\lambda_{N-1}} &=& \mu \frac{\d \lambda^R_{N-1}}{\d \mu} 
=0  +O(\A^2), 
\\
\beta_{\lambda_N} &=& \mu \frac{\d \lambda^R_N}{\d \mu} 
=0  +O(\A^2).
\end{eqnarray}
\end{mathletters}
Since there is no wavefunction nor diffusion constant renormalization
the set of one-loop RGEs can be summarized in the following way
\begin{equation}
\label{E:potRGE}
\mu \frac{\d V_R[\phi_R] }{\d \mu} 
= - {\A \over 8\pi \nu_R} V''_R[\phi_R]  +O(\A^2).
\end{equation}
This equation agrees with the computation of the RGEs based on the
effective potential, which was calculated in
Ref.~\cite{priority2c}. [See equation (51) of that paper.]
Furthermore, if we define $\mu = \mu_0 \exp(\tau)$, the previous RGE
becomes
\begin{equation}
\label{E:potRGE2}
\frac{\d V_R[\phi_R] }{\d \tau} 
= - {\A \over 8\pi \nu_R} {\d^2 V_R[\phi_R]\over \d\phi_R^2}  +O(\A^2),
\end{equation}
which implies the fact that the one-loop RGE in $d=2$ behaves like an
anti-diffusion process in field space.

At this point it is interesting to compare our results with
independent renormalization group results obtained, for example, by
Cardy in Ref.~\cite{Cardy}.  If a path integral is derived (along the
lines given in~\cite{Doi,Peliti,Grassberger}) for the two-body process
$A + A \rightarrow {\rm inert}$, then the corresponding RD equation
turns out to be given by~\cite{Cardy}
\begin{equation}\label{inert}
D\phi = -2\lambda \phi^2 + \eta(x),
\end{equation}
where however, the noise must be pure imaginary. A renormalization
group analysis of equation (\ref{inert}) shows that the field $\phi$
does not require wavefunction renormalization, nor does the diffusion
constant $\nu$ renormalize. Our one-loop heat kernel computation
performed for an {\it arbitrary} reaction polynomial, (85a,85b), is in
complete accord with these results (even though we treat real
noise). Returning to (\ref{inert}), the only non-vanishing
renormalization is that of the coupling $\lambda$. It turns out that
the one-loop renormalization group beta function is exact, and when
expressed in terms of the dimensionless renormalized coupling $g_R$ is
given by~\cite{Cardy}
\begin{equation}\label{beta-inert}
\beta(g_R) = b\,g_R^2,
\end{equation}
in $d=2$ dimensions, where $b$ is a positive constant (the value of
this constant is not specified in~\cite{Cardy}).

In order to compare these results, we define the following
dimensionless couplings
\begin{equation}
\label{E:dimensionless}
g_j \equiv \frac{\cal A}{8 \pi \nu_R}\, \frac{\lambda^R_{j+2}}{\lambda^R_j},
\qquad 0 \leq j \leq N-2,
\end{equation}
provided, of course, that for a given $j$ the coupling constant
$\lambda_j^R \neq 0$.  This definition together with the hierarchy of
beta functions given in (85) show that the dimensionless coupling
constants $g_j$ satisfy the following one-loop RGE
\begin{equation}
\label{E:dimlessRGE}
\beta(g_j) \equiv \mu \frac{\d g_j}{\d \mu} = g_j \, 
\left( \frac{\dot \lambda^R_{j+2}}{
\lambda^R_{j+2}} - \frac{\dot \lambda^R_{j}}{
\lambda^R_{j}} \right),
\end{equation}
where the overdot is shorthand notation for $\mu \, \d/{\d\mu}$.  We
now consider an RD equation of the type given in (\ref{E:RDD}) with
real noise and for a degree-two $(N=2)$ reaction polynomial
(\ref{E:potential-term}) $V[\phi] = \lambda_0 +
\frac{\lambda_2}{2}\phi^2$.  This particular choice is made in order
to be able to treat an RD equation as close as possible in structure
to the one in (\ref{inert}). Apart from the imaginary versus real
noise, the essential difference lies in the fact that we (must) have a
tadpole term, whereas (\ref{inert}) lacks such a term.  In this case
there is only one dimensionless coupling which can be defined, namely
\begin{equation}
g_0 = \frac{\cal A}{8\pi \nu_R}\, \frac{\lambda^R_2}{
\lambda^R_0},
\end{equation}
and equation (\ref{E:dimlessRGE}) implies the following one-loop beta
function for this dimensionless coupling
\begin{equation}
\beta(g_0) = g_0\Big(\frac{\dot \lambda_2^R}{\lambda_2^R} 
- \frac{\dot \lambda_0^R}
{\lambda_0^R} \Big) = -g_0\Big(\frac{\dot \lambda_0^R}{\lambda_0^R} \Big) = 
g^2_0.
\end{equation}
This follows from (85c) and from the fact that $\dot \lambda_2^R
\propto \lambda_4^R = 0.$

Thus, for the purposes of renormalization and calculating the RGEs,
this example demonstrates that it is equivalent to start from a
complex or real SPDE, and that the field theory can be derived from
microscopic principles or obtained by means of the procedure outlined
in~\cite{priority2a}.

\subsection{$d=3$ counterterms and renormalization}

In three space dimensions the divergent effective action is given by
equation (\ref{E:netaction})
\begin{eqnarray}
\Gamma_\epsilon^{(1-loop)}[\phi_R]&=& - 2 {\cal A} A_3\, \epsilon^{-1/4}
\int \d^3 \vec x \; \d t \; [c_{1}],
\end{eqnarray}
with $A_3={\Gamma(3/4)}/[{16 \pi (\pi \nu_R)^{3/2}}]$.

The net Seeley--DeWitt coefficient is given by
\begin{eqnarray}
\label{E:c-one-3}
[c_1]
&=&
V''[\phi_R](D\phi_R - V_R[{\phi_R}]).
\end{eqnarray}
The vanishing of the index one-half Seeley--DeWitt coefficient means
that the divergent effective action in $d=3$ becomes
\begin{eqnarray}
\Gamma_\epsilon^{(1-loop)}[\phi_R]&=& 
-{2}\A A_3\epsilon^{-1/4}
\int \d^3 \vec x \; \d t \; 
V''[\phi_R](D_R \phi_R - V[\phi_R]) +O(\A^2).
\end{eqnarray}
In order to determine the value of the counterterms we must once again
set
\begin{eqnarray}
\Gamma_\epsilon^{(1-loop)}[\phi_R]&=& - S_{\delta}[\phi_R]
{+ \rm finite}.
\end{eqnarray}
The last identification yields the following ($\mu$-independent) set
of counterterms
\begin{mathletters}
\begin{eqnarray}
\delta Z &=& 0  +O(\A^2),
\\
\delta \nu &=& 0  +O(\A^2), 
\\
\label{E:explicit-3d}
\delta \lambda_0 &=& -{\A A_3} \lambda^R_2  \epsilon^{-1/4}  +O(\A^2), 
\\
\delta \lambda_1 &=& -{\A A_3} \lambda^R_3  \epsilon^{-1/4}  +O(\A^2), 
\\
\vdots
\nonumber
 \\
\delta \lambda_{N-2} &=& 
-{\A A_3} \lambda^R_N  \epsilon^{-1/4}  +O(\A^2), 
\\
\delta \lambda_{N-1} &=& 0  +O(\A^2), 
\\
\delta \lambda_{N} &=& 0  +O(\A^2).
\end{eqnarray}
\end{mathletters}
As there is no scale dependent logarithmic divergence at one-loop
order in three-dimensions, all the beta functions
vanish~\cite{priority1a}.

\section{Discussion}

In this paper we have generalized and applied a method based on the
DeWitt--Schwinger proper time expansion to calculate the ultraviolet
divergences of the one-loop effective action associated to the RD
equation.  This particular approach involves the physical degrees of
freedom plus the ``ghost'' fields, which are needed to account for the
functional Jacobian that arises from a certain change of
variables~\cite{priority2a}. For RDs this Jacobian is generally
non-vanishing and must be taken into account in the computation of the
characteristic functional. The importance of the effective action lies
in the fact that it is the quantity needed to derive equations of
motion, which correctly take into account fluctuations and
interactions to a given number of loops.  The effective action encodes
the dynamics of the system. By contrast, the effective potential can
only tell us about static solutions. In order to know whether the
minima of the effective potential are relevant as solutions of the
late time behavior of the system we must see how accessible these
solutions are. But before any of these questions can be answered, the
effective action must be calculated.

The one-loop effective action is given in terms of a functional
determinant, which must be regulated and renormalized. The heat kernel
technique is an established method (used in QFT) for carrying this
out.  In QFT the functional determinant appearing in the one-loop
effective action is usually quadratic in both time $(\partial^2_t)$
and space derivatives $(\nabla^2)$.  In passing to a Euclidean
spacetime, the corresponding proper time equation for the kernel to be
solved (\ref{E:schrodinger}) is a heat equation for diffusion in
$n=d+1$ dimensions, with $s$ playing the r\^ole of diffusion
time. This is why its Green function (whether exactly or approximately
calculated) is justifiably denoted as the heat kernel. However, for
SPDEs, such as those considered in~\cite{priority2a}, the functional
determinant in the corresponding one-loop effective action involves
not only a ``mismatch'' between the number of spatial and temporal
derivatives, but also fourth- or even higher-order spatial
derivatives. We have seen an explicit example of this in the RD
equation treated here, which yields second order in time but fourth
order in space derivatives, $(\nabla^2)^2$, for its associated
operator determinant. (Even higher derivatives will be encountered in
the one-loop effective actions based on the Sivashinski and the
Swift-Hohenberg equations: two time derivatives but eight spatial
derivatives.) While much is known about the standard heat kernel and
its associated Seeley--DeWitt coefficients, very few (as far as the
authors are aware) of these ideas have been applied to other types of
field theories whose fluctuations may be of a non-quantum nature (\ie,
noise)~\cite{Gusynin}.  The heat kernel technique (and its
generalization) is well suited to regularize one-loop effective
actions, and therefore, is very useful to handle theories with
derivative-type interactions, as well as higher derivative ``kinetic''
terms.

We have applied these techniques to compute the one-loop effective
action for the RD equation.  As regards its ultraviolet
renormalizability, we found that the terms appearing at one-loop order
have the same structure, \ie, involve the same terms present already
at the level of the ``classical'' or zero-loop action.  Strictly
speaking, this claim holds true only if a certain bare constant is
added to the original equation of motion, as we have seen. This
constant, or ``tadpole'', is needed to carry out the one-loop
renormalization of the leading divergence that appears in the
effective action.  Moreover, this constant admits a simple physical
interpretation and can be ascribed either to a constant flux rate or
as the mean value of the additive noise source.  In regards to the
{}\emph{scale} of application of the RD equation, the one-loop
renormalizability indicates that although RD is a macroscopic
equation, [only intended to make physical sense for scales greater
than a certain minimum length scale $L_0$, defined by the underlying
molecular physics (if one is discussing chemical reactions)], we have
shown in this paper that the ultraviolet behavior of the RD equation
is controlled, and that considered as a strictly phenomenological
equation its short distance behavior is {\emph{much better than one
had any right to expect}}.  The short distance structure of this
effective action has been revealed via the calculation of the
Seeley--DeWitt coefficients up to one-loop order in the noise
amplitude. By means of this information, we have been able to
establish the one-loop finiteness of RD equations driven by additive
white noise for $d=0$ and $d=1$ space dimensions and their
renormalizability for $d=2$ and $d=3$ space dimensions.  In $d=2$
space dimensions there are logarithmic divergences which lead to
running, (scale dependence) of the parameters that describe the RD
equation.  There is no wavefunction renormalization at one-loop order.
These results hold for polynomial reaction kinetics of arbitrary
degree $N$. (Note: The absence of wavefunction renormalization has
already been demonstrated for the case of a (complex) RD equation
describing pair reactions ($N = 2$), where it turns out that the
one-loop beta function in $d=2$ is exact~\cite{Cardy}.)  When taken as
a model for pattern development, this result becomes even more
striking since this means we can safely use the RD equation to
investigate the important short distance and small time limit of the
field correlations that arise in pattern formation, as already
remarked earlier.

The RGE results obtained here are identical to those obtained (by
different means) from an effective potential calculation for RD
equations presented in Ref.~\cite{priority2c}. The effective potential
is the effective action restricted to constant field configurations
and plays an important r\^ole in uncovering patterns of symmetry
breaking and in the onset of pattern formation. Nevertheless, the
claim of the one-loop renormalizability made in~\cite{priority2c}
requires the investigation of the wavefunction renormalization which
was beyond the scope of that paper. The work presented here is also
intended to complete and complement that discussion.  Moreover, as
pointed out there, the combined effects of noise and interactions is
to shift the symmetric states of the system, as well as to change the
nature of the linear instabilities that may be induced by
perturbations around these new states. A spatial pattern is, by
definition, a spatially inhomogeneous configuration with a higher or
lesser degree of symmetry, if any such symmetry is at all present.
Thus, to investigate the onset of the spatial-temporal patterns
resulting from fluctuations and interactions, one must go beyond the
effective potential. This requires working with inhomogeneous fields
and the effective action.

The cautious reader will have noticed that most of the calculations
developed here depend solely on the reaction potential $V$ and its
derivatives and not on the fact that $V$ is a polynomial. In fact,
this is easy to see from the structure of our Seeley--DeWitt
coefficients, (\ref{E:c-half})--(\ref{E:c1}) and
(\ref{E:a-half})--(\ref{E:a1}).  Thus, the question arises if our
treatment can be extended and applied to handle RD equations with
non-polynomial reaction kinetics. The answer is in the affirmative
provided that the potential admits a real solution to (\ref{E:vev})
(\eg, $V[\phi] \sim \sinh[\phi]$) since we need a constant background
about which to expand the action, as indicated in
(\ref{E:netaction}). Provided such a solution exists, the rest of the
analysis presented here carries through as is, up to the extraction of
the renormalization group equations (which will again be non-vanishing
only for $d=2$). At this point the explicit functional form of the
potential changes the nature of the basis set of independent field
operators that leads to the RGEs.  Non-polynomial reaction kinetics do
indeed arise in many applications in chemical physics and in the
modelling of biological pattern formation, typically in the form of
rational functions (\ie, ratios of polynomials) and whenever
{\emph{constraints}} are to be imposed on the
model~\cite{Cross-Hohenberg,Walgraef}.  When a coarse-grained field
theoretic approach is applied to density waves in earthquakes for
example, a stochastic PDE for the (scalar) slip field (which measures
the relative displacement of two elastic media in contact taken along
the surface of contact) results which depends on the cosine of the
slip field, $\cos\phi$, and is driven by additive white
noise~\cite{Rundle}.  Thus, the work presented here is broad in scope.

Finally, these results also have the following practical application:
as analytic calculations can be carried only so far, it is clear that
numerical studies of SPDEs are crucial. Ultraviolet renormalizability
corresponds to the situation in which long distance physics is largely
insensitive to the details of short distance physics. In considering
numerical studies of the RD equation, we can therefore assert the
cut-off {\emph{insensitivity}} of the numerical solutions, at least to
one-loop. (In numerical studies, the ultraviolet cut-off is provided
by the lattice spacing.) This is of paramount importance since a
numerical study of RD will give us the information needed to see if
the system thermalizes, if it has steady state solutions, and most
importantly, if the minima of the effective potential calculated in
Ref.~\cite{priority2c} are explored in the time evolution of the
system.

\section*{Acknowledgements}

This work was supported in part by the Spanish Ministry of Education
and Culture and the Spanish Ministry of Defense (DH and CMP). In the
US, support was provided by the US Department of Energy (CMP and
MV). Additionally, MV wishes to acknowledge support from the Spanish
Ministry of Education and Culture through the Sabbatical Program, and
to recognize the kind hospitality provided by LAEFF (Madrid, Spain).
The authors would like to gratefully acknowledge discussions with Juan
\Perez--Mercader, whose interest and support was crucial in completing
this work.

\appendix  
\section{Free Green function for the RD effective action}
\label{A:free-green-function}

In this Appendix we calculate the free Green function appearing in
(\ref{E:formal}).  The only ``difficult'' part of the analysis is
dealing with the fourth-order spatial bi-harmonic operator
$(\nabla^2)^2$. Using translational invariance to set $x-x'\to x$, and
introducing Fourier transforms ($\d^n k = \d^d k\; d\omega$) as
follows
\begin{equation}
\label{E:FT1}
G_\free(\vec x t, \vec 0 0|s) = \int \frac{\d^n k}{(2\pi)^n}
\int_{-\infty}^{+\infty} \frac{\d\Omega}{2\pi} \;
\tilde G_\free(\vec k \omega|\Omega) \;
\exp [i(\vec k \cdot \vec x - \omega t - \Omega s)],
\end{equation}
we easily find that
\begin{equation}
\label{E:transform}
\tilde G_\free(\vec k\omega|\Omega) = \frac{1}
{\omega^2 + (\nu k^2)^2 - i \Omega},
\end{equation}
where $k^2 = {\vec k}\cdot \vec k$. By inverting the Fourier transform
(\ref{E:FT1}) one obtains the following integral representation for
the free Green function
\begin{equation}
G_\free(x,0|s) = \int \frac{\d^dk}{(2\pi)^d}\int 
\frac{\d\omega}{2\pi}
\int \frac{\d\Omega}{2\pi}\;
\frac{1}
{\omega^2 + (\nu k^2)^2 - i \Omega} \;
\exp[{i(\vec k\cdot \vec x -\omega t -\Omega s)}].
\end{equation}
We first perform the integral over $\Omega$ by means of a contour
integral in the complex $\Omega$-plane.  There is only one simple pole
on the negative imaginary $\Omega$-axis and we close the
(semi-circular) contour (centered at the origin) in the lower half
plane. As the radius of this arc goes to infinity, only the
contribution along the real $\Omega$-axis remains ($s > 0$).
As a result and by the Residue Theorem we have (the contour is closed in
the clockwise sense)
\begin{equation}
\label{E:result}
G_\free(x,0|s) = \int \frac{\d^dk}{(2\pi)^d}\int \frac{\d
\omega}{2\pi} \;
\exp [{i(\vec k\cdot \vec x -\omega t)}] \;
\exp( -s [\omega^2 + (\nu k^2)^2 ]) . 
\end{equation}
We can go further and compute the $\omega$ integral exactly to obtain
\begin{equation}
G_\free(x,0|s) = \left(\frac{1}{4 \pi s}\right)^{\half}
\exp \left( {-\frac{ t^2}{4s}} \right)
\int \frac{\d^dk}{(2\pi)^d}\;
\exp\left[ {-s(\nu k^2)^2 + i\vec k
\cdot \vec x} \right],
\end{equation}
and the remaining momentum integral is manifestly convergent (for $d >
0$). As for the boundary condition, note that for $s \rightarrow 0$,
$\lim_{s \rightarrow 0} G^{({\mathrm free})}(\vec x t,\vec 0 0 |s) =
\delta(t,0)\delta^d(\vec x,\vec 0)$, (in the sense of distributions)
as it must, since
\begin{eqnarray}
\delta(t) \sim 
\lim_{s \rightarrow 0} \left( \frac{1}{4 \pi s} \right)^{\half}
 \exp[-t^2/(4s)],
\end{eqnarray}
and
\begin{eqnarray}
\delta^d(\vec x) 
= 
\int \frac{\d^d \vec k}{(2\pi)^d} \; e^{i\vec k \cdot \vec x}.
\end{eqnarray}
One can also check that the boundary condition is satisfied, before
integrating over $\omega$, by simply setting $s=0$ in
(\ref{E:result}).  Let us now work out the momentum integration.

The angular integration is given by
\begin{equation}
\int \d\Omega_{d-1} \exp\left(i\vec k\cdot \vec x \right)
=
(2\pi)^{d/2} \;
{J_{(d-2)/2}(kx) \over {(kx)}^{(d-2)/2}}.
\end{equation}
The exact Green function or kernel for our free ``heat'' operator in
(\ref{E:freekernel-1}) is
\begin{equation}
\label{E:green2}
G_\free(\vec xt,\vec 0 0|s) 
= \left(\frac{1}{4 \pi s}\right)^{\half}
\exp \left({-\frac{t^2}{4s}} \right)
\frac{1}{(2\pi)^{d/2}}\;
\int_0^{+\infty} k^{d-1}\d k \; 
\exp[ - s(\nu k^2)^2 ]
\left[ \frac{J_{(d-2)/2} (k |x|)}{ (k|x|)^{(d-2)/2} }\right].
\end{equation}
This solves the differential equation (\ref{E:freekernel-1}) and
satisfies the boundary condition $G_\free(\vec x,t|0) = \delta^d(\vec
x) \;\delta(t)$ for vanishing proper time $s$, hence this is the {\em
unique} solution of (\ref{E:freekernel-1}).

\noindent
\emph{Important point:} as remarked above, translational invariance
implies that $G_\free(\vec x t,\vec y t'|s) = G_\free(|\vec x -\vec
y|,(t- t')|s)$. We are treating stochastic processes on a flat
$d+2$-dimensional background ($d$-space dimensions plus real physical
time $t$ plus Schwinger proper time $s$).

Using the Taylor series representation
\begin{equation}
\frac{J_\ell(z)}{z^\ell} = \left(\frac{1}{2}\right)^{\ell}
\sum_{n=0}^{+\infty} \frac{(-)^n \; (z/2)^{2n}}{n! \; \Gamma(\ell+n+1)},
\end{equation}
and integrating (\ref{E:green2}) term-by-term, we find that (after
making use of the time and space translational invariance of the Green
function)
\begin{eqnarray}
\label{E:master}
G_\free(\vec x t,\vec x' t'|s) 
&=& 
\left(\frac{1}{4 \pi s}\right)^{\half} \;
(s \nu^2)^{-\frac{d}{4}} \;
\exp \left[{-\frac{(t-t')^2}{4s}} \right] \;
\frac{1}{2^{d+1}\pi^{d/2}} 
\sum_{n=0}^{+\infty} \left(-\frac{1}{4}\right)^n \; 
\frac{\Gamma(\frac{n}{2} + \frac{d}{4})}{n! \; \Gamma({d\over2}+n) } 
\left( \frac{|\vec x-\vec x'|^2}{\nu \sqrt{s}} \right)^n 
\\
&=& 
A_d 
\;
\exp \left[{-\frac{(t - t')^2}{4s}} \right] \;
s^{-\half - \frac{d}{4}} \;
\sum_{n=0}^{+\infty} \left(\frac{-1}{4}\right)^n \; 
C_{d,n} \; \frac{ \Gamma({d\over2}) }{ n! \; \Gamma({d\over2}+n ) } 
\left( \frac{|\vec x-\vec x'|^2}{\nu \sqrt{s}} \right)^n ,
\end{eqnarray}
where 
\begin{equation}
A_d = 
\left(\frac{1}{4\pi}\right)^{\half} \;
\frac{\pi^{\frac{d}{2}} \; \Gamma(\frac{d}{4}) }
{2(2\pi)^d \; \Gamma({d\over2})}
\left( \frac{1}{\nu^2} \right)^{\frac{d}{4}}
, 
\;\;\; {\rm and}\;\;\;
C_{d,n} = {\Gamma({d+2n\over4})\over\Gamma({d\over4})}.
\label{E:a-cde}
\end{equation}
Special attention should be paid to the fact that this free Green
function involves a series in {}\emph{half-integer} powers of
Schwinger proper time, $\sqrt{s}$, a feature that we use in choosing
our ansatz for the full Green function.

\section{Homogeneous field configurations}
\label{A:homogeneous}

For constant (homogeneous and static) fields, $\phi=v_0$, the
associated ``heat kernel'' can be solved for {}\emph{exactly}.  We
simply present the final result and skip the details of a calculation
that is an analog of that for $G_{\free}(x,x'|s)$.  The on-diagonal
Green function $G^{(0)}_\field(x,x|s)$ is given by
\begin{eqnarray}
\label{E:constG-coincidence3}
G^{(0)}_\field(x,x|s) 
&=& 
A_d \; s^{-\half-{d\over4}} \;
\exp \left [ -s(V''[v_0]V[v_0] + V'[v_0]V'[v_0])\right] 
\sum_{\ell=0}^{+\infty} (\sqrt{s} \; 2 \; V'[v_0] )^l \; 
{C_{d,\ell}\over \ell!}
\nonumber \\
&=& G_{\free}(x,x|s) \;
\exp \left[-s(V''[v_0]V[v_0] + V'[v_0]V'[v_0])\right] 
\sum_{\ell=0}^{+\infty} (\sqrt{s} \; 2 \; V'([v_0] )^l \; 
{C_{d,\ell}\over \ell!},
\end{eqnarray}
where we have again made use of the definition $C_{d,\ell}$.  We
conclude that
\begin{equation}
\sum_{\ell=0}^{+\infty} [b_{\frac{\ell}{2}}] \; s^{\ell/2} =
\exp \left[-s \left ( V''[v_0]V[v_0] + V'[v_0]V'[v_0] \right) \right]
\sum_{\ell=0}^{+\infty} (\sqrt{s} \; 2 \; V'[v_0] )^l \;  
{C_{d,\ell}\over \ell!}.
\end{equation}
We now match the first fractional powers in $s$ and obtain the
Seeley--DeWitt coefficients for a constant field configuration $v_0$
\begin{eqnarray}
[b_0] &=& 1,
\\
\label{E:b-half-homogeneous}
{}[b_{\half}]
&=& 2 \; C_{d,1} \; V'[v_0],
\\
\label{E:b1-homogeneous}
{}[b_1] 
&=&  
-  V''[v_0]\;V[v_0].
\end{eqnarray}
The coefficients for $G_\ghost^{(0)}$ can be immediately obtained from
the previous Seeley--DeWitt coefficients by setting $V''[v_0] =
0$. That is
\begin{eqnarray}
[a_0] &=& 1,
\\
\label{E:a-half-homogeneous}
{}[a_{\half}] &=& 2 \; C_{d,1} \; V'[v_0],
\\
\label{E:a1-homogeneous}
{}[a_1] &=& 
\left({d\over 2} - 1\right)\;V'[v_0]\;V'[v_0].
\end{eqnarray}
Finally, for the net Seeley--DeWitt coefficients, we can write
\begin{eqnarray}
[c_0] &=& 0,
\\
\label{E:c-half-homogeneous}
{}[c_{\half}] &=& 0,
\\
\label{E:c1-homogeneous}
{}[c_1] &=& -  V''[v_0]\;V[v_0].
\end{eqnarray}
These results are consistent with the net integrated Seeley--DeWitt
coefficients and with the physical and ghost integrated Seeley--DeWitt
coefficients presented in the body of the paper [equations
(\ref{E:c-half})--(\ref{E:b1})].

\section{The Feynman--Hellman formula}
\label{A:feynman-hellman}

We write the Feynman--Hellman formula in the
form~\cite{Feynman-Hellman}
\begin{equation}
{\d\over\d\e} \exp (A+\e B)
=
\int_0^1 \d\ell \; 
\exp[ \ell(A+\e B)] \; B  \; \exp[(1-\ell)(A+\e B)].
\end{equation}
This equation is central to the computation of the Seeley--DeWitt
coefficients presented in the paper. For instance, if we take the
trace and then use the cyclic property, we can write
\begin{equation}
{\d\over\d\e} \Tr [\exp(A+\e B)]
=
\Tr\left\{ B \; \exp[  (A+\e B)] \right\}.
\end{equation}
We differentiate the previous equation to obtain
\begin{equation}
{\d^2\over\d\e^2} \Tr [ \exp(A+\e B)]
=
\int_0^1 \d\ell \;
\Tr\left\{ B \; \exp[ \ell(A+\e B)]  \; B  \; \exp[(1-\ell)(A+\e B)]  \right\}.
\end{equation}
We can conclude that
\begin{eqnarray}
\Tr [ \exp(A+\e B)]
&=& 
\Tr[\exp(A)]  + 
\e \Tr [ B \; \exp(A)] 
+
{\e^2\over2} \int_0^1 \d\ell \;
\Tr\left\{ B \; \exp (\ell A)  \; B  \; \exp[(1-\ell)A])\right\} 
 +
O(\e^3).
\end{eqnarray}
This perturbative expansion is the basis for extracting the first two
{}\emph{integrated} Seeley--DeWitt coefficients.



\begin{thebibliography}{99}
\bibitem{QFT} 
There is a wealth of excellent monographs and texts treating the
subject of quantum field theory. A small and select sampling of some
modern expositions may be found in~\cite{Ramond,Zinn,Rivers,Weinberg}.
For more additional references, see the bibliography
in Ref.~\cite{phi4}.
\end{thebibliography}
\end{document}